\documentstyle[12pt,aaspp4]{article}

\def\etal{{et al.}}
\def\asca{{\it ASCA}}

\def\exosat{{\it EXOSAT}}

\def\ros{{\it ROSAT}}

\newbox\grsign \setbox\grsign=\hbox{$>$} 
\newdimen\grdimen \grdimen=\ht\grsign
\newbox\simlessbox \newbox\simgreatbox \newbox\simpropbox
\setbox\simgreatbox=\hbox{\raise.5ex\hbox{$>$}\llap
     {\lower.5ex\hbox{$\sim$}}}\ht1=\grdimen\dp1=0pt
\setbox\simlessbox=\hbox{\raise.5ex\hbox{$<$}\llap
     {\lower.5ex\hbox{$\sim$}}}\ht2=\grdimen\dp2=0pt
\setbox\simpropbox=\hbox{\raise.5ex\hbox{$\propto$}\llap
     {\lower.5ex\hbox{$\sim$}}}\ht2=\grdimen\dp2=0pt

\begin{document}

\title{On X-ray Variability in Seyfert Galaxies}

\author {T.J.Turner \altaffilmark{1,2}, 
I.M. George \altaffilmark{1, 3}, 
K. Nandra \altaffilmark{1, 3} 
D. Turcan \altaffilmark{4}
}

\altaffiltext{1}{Laboratory for High Energy Astrophysics, Code 660,
	NASA/Goddard Space Flight Center,
  	Greenbelt, MD 20771}
\altaffiltext{2}{University of Maryland Baltimore County, 1000 Hilltop Circle,
	Baltimore, MD 21250}
\altaffiltext{3}{Universities Space Research Association}
\altaffiltext{4}{University of Maryland College Park}

\slugcomment{To be submitted to {\em The Astrophysical Journal}}

\begin{abstract}

This paper presents a quantification of the X-ray variability 
amplitude for 79 \asca\ observations of 36 Seyfert 1 galaxies. 
We find that consideration of sources 
with the narrowest permitted lines in the optical band introduces 
scatter into the established 
correlation between X-ray variability and nuclear luminosity. 
Consideration of the X-ray spectral index and variability properties 
together shows distinct 
groupings in parameter space for broad and narrow-line Seyfert 1 galaxies, 
confirming previous studies.
A strong correlation is found between hard X-ray 
variability and FWHM H$\beta$. A range of nuclear mass and accretion rate  
across the Seyfert population can 
explain the differences observed in X-ray and optical properties. 
An attractive 
alternative model, which does not depend on any systematic 
difference in central mass, 
is that the circumnuclear gas of NLSy1s is 
different to BLSy1s in temperature, optical depth, density or 
geometry.

\end{abstract} 


\section{Introduction}

\label{sec:intro} 

X-ray flux variability has long been known to be a common property of active
galactic nuclei (AGN). Ariel {\sc V} and HEAO-1 first revealed long term (days to
years) variability in AGN  (e.g. Marshall, Warwick \& Pounds 1981). It was not
until the long-duration and uninterrupted observations of {\it EXOSAT} that
rapid (thousands of seconds) variability was also established as common in
these sources. X-ray flux variations are observed on timescales from 
$\sim$ a 
thousand seconds to years and amplitude variations of up to an order of
magnitude are observed in the $\sim$ 0.1-10 keV band (also see 
review by Turner 1992 and references therein). 
Barr \& Mushotzky (1986) suggested 
the flux-doubling timescale of an AGN to be inversely proportional to
its luminosity, both properties measured in the hard X-ray regime. 
The use of ``doubling timescale'' was a somewhat unsatisfactory
quantification of the X-ray variability as, for many sources, this had to be
derived by extrapolation of lower amplitude events.
Lawrence \& Papadakis  (1993) confirmed the variability-luminosity 
relationship, quantifying  variations seen in 
\exosat\ light curves by using the timescale above which the integrated 
rms variation should be 10\%. Green, McHardy and Lehto (1993) also confirmed 
the correlation using the ``normalized variability amplitude'' and further, 
suggested 
sources with steeper X-ray spectra showed the highest amplitude of variability, 
perhaps due to a lower contribution from a reprocessed component. 
The existence of a correlation between X-ray variability and 
spectral slope was supported by the work 
of Koenig, Staubert \& Wilms (1997), using a different 
quantification of X-ray flux variability. 
Boller \etal\ (1996) examined 
the \ros\ ``variation timescale''  as a function of luminosity, 
concluding the high amplitude of variability observed in 
narrow-line Seyfert 1 galaxies to be 
 inconsistent with scattering in an extended region, like that 
suggested for Seyfert 2 galaxies.

Nandra \etal\ 
1997a (N97a) initiated use of
``excess-variance'' , $\sigma^{2}_{\rm RMS}$ to quantify the X-ray 
variability of 
AGN (see below). This quantity had been previously used in analysis
of the ultraviolet time series of AGN where Edelson, Krolik \& Pike (1990) 
found an anticorrelation between excess-variance and 
luminosity.  Using excess-variance, N97a found 
an inverse correlation between X-ray variability (sampled using
128 s bins) and luminosity for a sample of Seyfert 1 galaxies 
with predominantly broad permitted lines. 
Fiore et al. (1998) used {\it ROSAT} HRI observations of a sample of quasars 
to examine correlations between the (soft) X-ray parameters of 
excess-variance, spectral index and luminosity, finding 
steep-spectrum, narrow-line quasars to show larger amplitude 
of X-ray variability 
in the 0.1 -- 2 keV band than the flat-spectrum broad-line quasars.

We recognize that there is a continuous distribution of optical line widths 
in Seyfert 1 galaxies, and any attempt to split these into narrow-line and 
broad-line Seyfert 1s is arbitrary. However, 
for convenience we do make such a 
split here by referring to objects with FWHM H$\beta < 2000$ km/s 
as narrow-line Seyfert 1s (NLSy1s) and objects with FWHM H$\beta > 2000$ km/s
as broad-line Seyfert 1s (BLSy1s). This is purely to facilitate 
a concise discussion of the different regimes within the
Seyfert 1 distribution. 

The NLSy1 Ton S180 was found to lie 
significantly above the correlation established for the N97a sample (Turner
\etal\ 1998) and also showed energy-dependent variability. 
The TON S180 result 
indicated  that analysis of a sample including a significant number 
of NLSy1s may 
yield further insight into the physical processes at work in AGN. 

This paper presents a timing analysis of 79 observations of 
36 Seyfert 1 galaxies. 
Our sample consists of Seyfert 1 galaxies available from the 
Advanced Satellite for Cosmology and Astrophysics  
({\it ASCA}) archive up to November 1998. 
The purpose of this investigation was to expand upon the X-ray timing results
presented in N97a, and to search for confirmation of some correlations 
between hard X-ray and optical 
parameters, suggested by previous works (above).  
In particular, we aimed to determine whether the
established luminosity-variability correlation would alter when 
Seyfert galaxies 
covering the full observed range of permitted line widths 
were considered. 
The objects presented here do not form a complete sample.
Nevertheless we suggest the  results from this large collection of
objects provides important new insight into the AGN phenomenon.

\section{ASCA Observations and Data Reduction}

{\it ASCA} has two solid-state imaging spectrometers 
(SISs; Burke \etal\ 1994) and two gas imaging spectrometers 
(GISs; Ohashi \etal\ 1996) yielding data over 
an effective bandpasses $\sim$0.5--10~keV. 
The data presented here were systematically 
reduced in the same way as the Seyfert galaxies presented in N97a, 
using source events within 
extraction cells of radii 4.8 and 6.6 arcmin for the SIS and
GIS data, respectively. 

In the timing analysis we used only  data from the SIS instruments. We
required that time series included in our analysis have at least 20 counts per
time bin, and at least 20 bins in the final light curve. To increase the
signal-to-noise ratio, we combined the SIS0 and SIS1 detectors requiring all
time bins to be at least 99\% exposed. The background level was not subtracted
from these light curves. The excess-variance quantifies 
amplitude of variability in excess of that expected from statistical 
fluctuations in the background level. For each dataset we verified 
that the background light curve did not contain any significant trends. 
N97a use a
128s time bin for the light curve analysis. However, many of our AGN are
faint, and analysis using a 128 s time bin would have lead to the exclusion of
many potentially interesting sources. We found 256 s to be the optimum bin
size for inclusion of most of the available data (use of longer integration
times would have resulted in too few bins for many datasets).

As noted above, N97a introduced the use of the quantity 
normalized `excess-variance, $\sigma^{2}_{\rm RMS}$ 
for analysis of \asca\ time series.  Following that paper
we designate the count rates for the $N$ points in each light curve 
as $X_{\rm i}$, with errors 
$\sigma_{\rm i}$. We further define $\mu$ as the unweighted,
arithmetic mean of the $X_{\rm i}$. Then:

$$ \sigma^{2}_{\rm RMS}=\frac{1}{N\mu^{2}}\sum_{i=1}^{N} [(X_{\rm
i}-\mu)^2-\sigma_{\rm i}^{2}] $$

The error on $\sigma^{2}_{\rm RMS}$, is given 
by $s_{\rm D}/(\mu^2\sqrt{N})$ where: 
$$  
s_{\rm D}^{2}=\frac{1}{N-1}\sum_{i=1}^{N} 
\Big\{ [(X_i-\mu)^2-\sigma_{i}^{2}]-\sigma^{2}_{\rm RMS}\mu^2 \Big\}^2
$$  

i.e. the variance of the quantity $(X_i-\mu)^2-\sigma_{i}^{2}$. 
(Note there was a typographical error in N97a in that the equation 
for the error on 
$\sigma^{2}_{\rm RMS}$ should have had the quantity inside the summation 
squared, as shown here). 
We refer the reader to N97a for a discussion of the merits of
using this quantity for timing analysis.

We note that for perfect comparison between sources, the excess-variance
should be calculated using observations of the same duration ($T_D$). 
However, in 
practice this would limit the analysis by reducing the number of sequences we
could consider, or would require truncation of light curves to match the
shortest observation. We note many of our light curves were constructed from
observations of duration close to 50 ks. Our analysis includes observations
both a factor of 2 smaller and larger than this. As the duration of an
observation is random with respect to the parameters of interest then the
effect of allowing a range of observation durations is to introduce scatter,
which could hide some weak correlations but which should not introduce false
correlations (because $T_D$ is not correlated with $\Gamma$, luminosity or
FWHM H$\beta$).

\section{Results}

We recalculated $\sigma^{2}_{\rm RMS}$ for the N97a sample of objects 
and all the new data presented here using 256 s time bins. 
Table~1 shows the datasets used in our analysis. 
The following sequences were also analyzed, but failed to meet our 
criteria for counts per bin and/or  number of bins in the final light curve:
Fairall-9 (73011050); NGC 4151 (70000000, 70000010, 71019030, 71019000); 
NGC 6814 (70012000); Mrk 509 (74024000, 74024010, 74024020); NGC 7469 
(71028000, 71028030, 71028010); MCG --2-28-22 (70004000); ESO 141-G55 
(72026000); PG1211$+143$ (70025000); Mrk 957(75057000) and Mrk 507(74033000). 

The results are shown in Figure~1a. The dark squares denote 
NLSy1s, BLSy1s 
are denoted by a simple cross. Error bars are $1 \sigma$. 
Compared to the N97a sample, we have most notably added a 
significant 
number of objects with relatively narrow permitted lines  and 
a number of points at the high luminosity end of the 
distribution (with multiple observations of 
Mrk 509 and Fairall-9).  

Examination of Figure 1a shows that there remains a significant 
anticorrelation between $\sigma^{2}_{\rm RMS}$ and luminosity 
(a Spearman-Rank correlation coefficient of -0.683 is obtained, 
 significant at $> 99$\% confidence). However there is a 
much greater degree of scatter in the correlation  
than in that shown by N97a. It is evident that  this 
is predominantly due to the inclusion of a relatively large number of 
NLSy1 objects. 
This was expected, based upon the Ton S180 result 
(Turner \etal\ 1998) and  was 
suggested by the \ros\ results of Boller \etal\ (1996). 
This result has also been found in a large independent study of NLSy1s 
(Leighly et al. 1999). 
This scatter suggests that $\sigma^{2}_{\rm RMS}$ 
has a strong dependence on FWHM H$\beta$ in addition to the dependence 
on luminosity. This does not weaken the dependence of $\sigma^{2}_{\rm RMS}$
on luminosity, in the sense that it appears that  objects  with 
similar values of 
FWHM H$\beta$ still show a correlation between $\sigma^{2}_{\rm RMS}$ 
and luminosity. 

NLSy1s are known to be steep, even in the {\it ASCA} bandpass
 (Brandt \etal\ 1997), and {\it ROSAT} data showed them to be 
generally highly variable X-ray sources (e.g. Boller \etal\ 1996, 1997;
Forster \& Halpern 1996). 
Work by Green \etal\ (1993) showed  a relationship 
between X-ray index and ``normalized variability amplitude'' 
in a sample of \exosat\ observations of AGN. The Green et al (1993) 
result was supported by the results  of Koenig et al (1997). 
Hence  we examined the relationship between $\sigma^{2}_{\rm RMS}$ and 
the 2-10 keV  photon index $\Gamma_{2-10}$.
The photon indices ( Table~1) were determined in the (rest-frame) 2--10 keV 
band for each source, excluding the (rest-frame)  5-7.5 keV band  which 
can contain significant photons from an Fe K$\alpha$ line (confusing 
measurements of the hard X-ray index). The photon index was derived fitting 
 the two SIS and GIS datasets simultaneously, and using background 
spectra extracted from source-free regions of the detectors. 
A small difference in the normalization of each instrument was allowed, to 
account for slight differences in the flux calibration of each ($< 20$\%). 
Absorption was included in the spectral model, but the column 
densities were allowed 
to be relatively unconstrained (simply greater than or equal to the Galactic 
line-of-sight value in each case). This resulted in conservative error bars for
$\Gamma$. Column densities were generally consistent with Galactic 
values and are not tabulated here. Some values of $\Gamma_{2-10}$ were 
taken from N97b and George \etal\ (1998a), as noted in Table~1. 
These quantities are plotted in Figure 1b, and the correlation yielded 
a Spearman Rank coefficient of 0.412. While the correlation is 
significant at $> 99$\% confidence, the figure  could be interpreted 
as showing that NLSy1 and BLSy1 objects occupy different areas 
of parameter space, with some zones-of-avoidance. Sources with steep 
photon indices, 
$\Gamma_{2-10}$ show relatively large amplitudes of variability 
when sampled on a timescale of 256s, compared to 
flat-spectrum objects. 

Next we examined directly the relationship between FWHM H$\beta$ 
and $\sigma^{2}_{\rm RMS}$. This has not previously been done 
in a quantitative manner using \asca\ data. 
Values of FWHM H$\beta$  were extracted from the 
literature and are shown in Table~1. 
We caution the reader that many sources are known to show significant 
variations in FWHM H$\beta$  over 
timescales of years, and we take just one representative value 
for each source. 
In particular, Mrk~290 has reported values for FWHM H$\beta$ 
which vary a great deal, we have used the Osterbrock (1977)  value 
but note a much wider line of FWHM H$\beta=5340$ km/s has also been reported 
(Boroson \& Green, 1992).  
(We also caution that the least reliable numbers in Table~1 
are the FWHM H$\beta$ 
 for EXO 055620-3 and Nab 0205$+024$, which we measured from 
published optical spectra, but we deemed these estimations adequate for the 
purpose of this paper). 

Figure~1c shows a very strong  correlation between FWHM H$\beta$ and
$\sigma^{2}_{\rm RMS}$. The Spearman-Rank correlation coefficient was
-0.723, significant at $> 99$\% confidence, and we find
 $\sigma^{2}_{\rm RMS} \propto$ (FWHM H$\beta)^{-2.8}$. It is remarkable that 
this correlation should appear so strong given the line width 
variability noted above, and the lack of simultaneity of the optical and 
X-ray measurements. There must be a very strong underlying 
link between these two observables, which we will discuss in detail later.

As there is a known correlation between EW H$\beta$ and FWHM H$\beta$ (Boroson
\& Green 1992), one would expect to also see a correlation of $\sigma^{2}_{\rm
RMS}$ with EW H$\beta$. Indeed these quantities appear linked (Figure 1d) with
a Spearman Rank correlation coefficient of -0.523 (significant at $> 99$\%
confidence). However, as we have relatively few measurements 
available for EW H$\beta$, we concentrate on the correlation 
between excess-variance and FWHM H$\beta$. 

It is also evident that 
significant changes are observed in  $\sigma^{2}_{\rm RMS}$ between 
X-ray  observations of a single source, although  only Fairall-9, NGC 3783, NGC 5548 and 
Mrk 509 were observed frequently enough to look for this. 
In fact N97a previously  noted changes 
in $\sigma^{2}_{\rm RMS}$ for NGC 4151,  
and George \etal\ 1998(b) noted them for NGC 3783. As the observations 
of any particular source are generally of 
approximately the same duration,  these results indicate that the 
process producing the variability is non-stationary (Table~1). 

\section{Discussion}

Assuming the bulk motion of the broad-line-gas is virialized, 
then differences in line width can be attributed 
to the emitting  gas existing further from the nuclei of  
 NLSy1s than  BLSy1s; alternatively 
the gas is at the same radius in both but the 
mass of the central black hole is smaller in NLSy1s than in BLSy1s. 
Our strongest correlation appears between $\sigma^{2}_{\rm RMS}$ and 
FWHM H$\beta$, and is consistent with rapid variability and 
narrow lines being a result of small central mass. 
The X-ray luminosities show comparable distributions for NLSy1s 
and BLSy1s. 
Taking a BLSy1 and NLSy1 of the same X-ray luminosity, 
then the BLSy1 will be a relatively high-mass system and the NLSy1 
will be a low-mass system radiating at  a high fraction of 
its Eddington luminosity 
to achive the same luminosity (yet having more rapid variability for 
that luminosity). 
These simple arguments have been discussed by numerous 
authors (e.g. Laor et al 1997).  
When both extremes of the Seyfert 1 phenomenon are 
considered in a variability-luminosity  analysis then it includes 
sources with a range of accretion rates 
producing the same luminosity, this naturally introduces 
significant scatter into the correlation. 

We confirm the findings of Green et al (1993) and  
Koenig et al (1997), that 
Seyferts with flat X-ray spectra (in the 2-10 keV band) 
show relatively low amplitude of 
X-ray variability (when sampled on 256s) compared to the 
steep sources. 
The relatively high accretion rate inferred for NLSy1s 
may result in different conditions in the 
accretion disk and its corona. A higher optical depth of the corona 
in BLSy1s (compared to NLSy1s) 
 would produce the harder X-ray spectrum observed. 
Another key parameter is coronal temperature, systematically 
cooler coronae could exist in NLSy1s, producing few hard X-ray photons. 
The size of the Comptonization region is also important, 
an extended region will smear the intrinsic nuclear variability. 
Similarly, the geometry of the Comptonizing gas is important, relative 
to our line-of-sight.

A serious concern for models based upon a fundamental difference 
in central mass  is that Rodriguez-Pascual \etal\ 
(1997) find broad components to emission lines observed in the UV band of both 
NLSy1s and BLSy1s. Again, assuming this gas is virialized it then  contradicts
 the hypothesis of systematically smaller masses in NLSy1s than BLSy1s. 
We consider whether there is a model to explain the observed 
effects without relying on a relatively small central mass for NLSy1s. 
One clue may come from the  
observation of energy-dependent variability in 
TON S180 (Turner \etal\ 1998) and some BLSy1s 
(N97a, George \etal\ 1998a,c). It appears that in those sources 
the hard X-rays have traversed a longer path than soft 
X-ray photons, probably due to Compton-upscattering of some
fraction of the nuclear continuum. 
Thus differences in the coronal conditions {\it alone} could explain 
both the difference in X-ray variability and spectral properties between 
BLSy1s and NLSy1s, 
with no inference on central mass. The problem with this picture is 
that the origin of the coronal differences is unknown, and our 
data require differences in coronal conditions to lead to 
differences in observed FWHM H$\beta$. 
Rodriguez-Pascual \etal\ (1997) have suggested that the line-emitting 
material is optically thin in NLSy1s, and that observed properties 
of any Seyfert depend 
on the gradient of density and ionization-state of the circumnuclear 
material around the Seyfert nuclei. The extension of this to our 
results indicates differences in nuclear environment, rather than central mass, 
can explain the range of observational properties of each type of
Seyfert galaxy. 

Some previous works have explained the narrow  widths of optical 
lines in NLSy1s by obscuration of the inner region
of high-velocity gas evident in BLSy1s. This model now seems 
unlikely, since X-ray observations have 
revealed little or no absorption of the nuclei of NLSy1s (e.g. 
Boller \etal\ 1996, Comastri \etal\ 1998, Turner \etal\ 1998). 

\section{Conclusions}

The correlation established previously between X-ray variability and nuclear
luminosity contains significant scatter when NLSy1 galaxies are included in
the analysis, this result is in agreement with that found in the independent
study of Leighly et al (1999). A strong correlation is also found between
variability and FWHM H$\beta$. A range of nuclear mass and accretion rate can
explain differences in X-ray variability, X-ray spectrum and optical-line-widths
across the Seyfert population. However, evidence from UV data suggests we must
consider models which do not hinge on differences in central mass. An
alternative model is that the environment of 
the nuclei of NLSy1s is significantly
different to BLSy1s. Differences in temperature, optical depth, density and
geometry of circumnuclear gas can explain all observational results.

\vspace{2cm}
We are grateful to the {\it ASCA} team for their operation of the satellite. 
 This research has 
made use of the NASA/IPAC Extragalactic database,
which is operated by the Jet Propulsion Laboratory, Caltech, under
contract with NASA; of the Simbad database, 
operated at CDS, Strasbourg, France; and data obtained through the 
HEASARC on-line service, provided by NASA/GSFC. 
This work was supported by NASA Long Term Space Astrophysics grant 
NAG 5-7385.

\pagestyle{empty}
\setcounter{figure}{0}
\begin{figure}
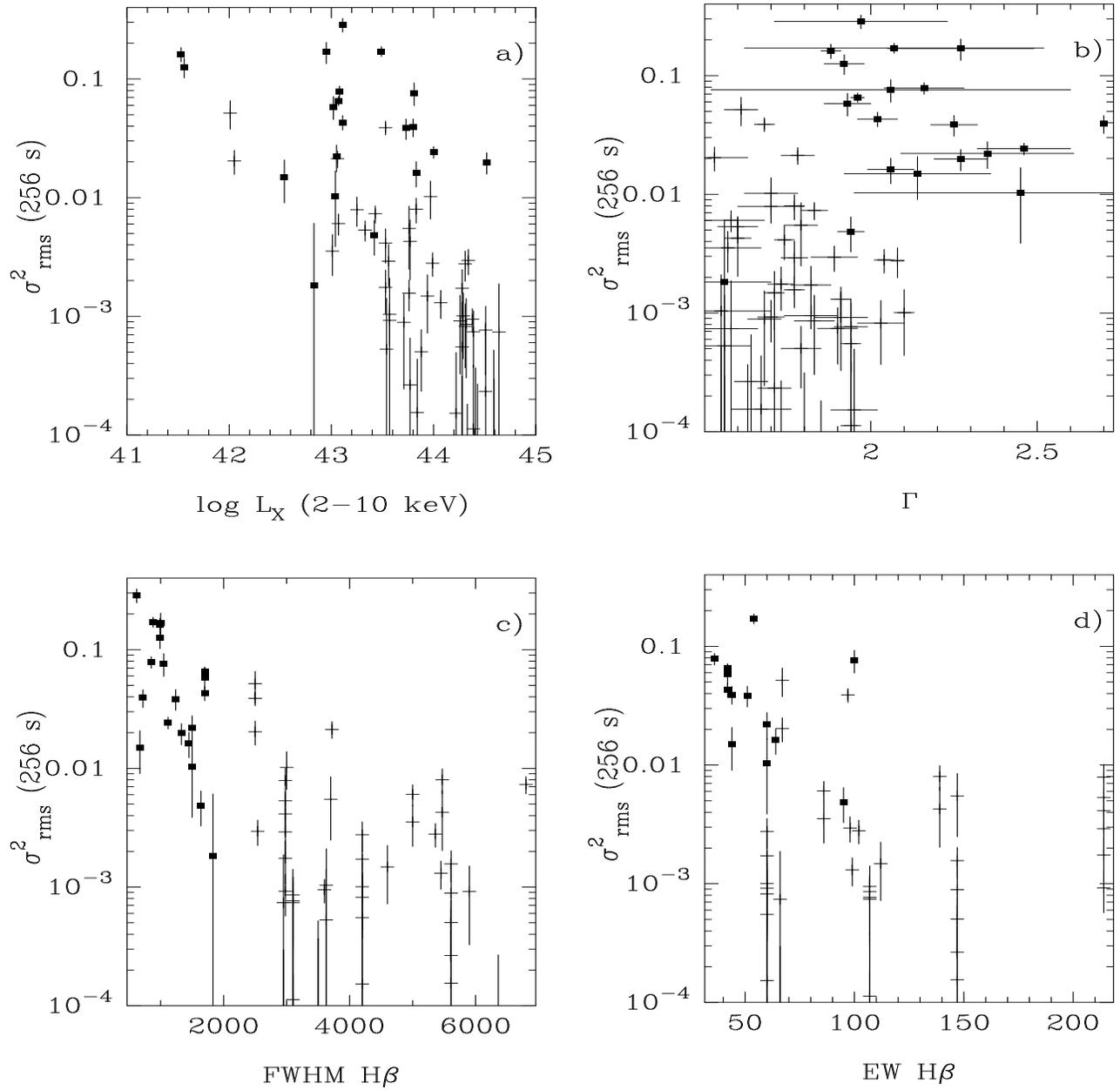

\thispagestyle{empty}
\plotfiddle{panel1.vps}{2cm}{0}{45}{35}{-250}{0}
\plotfiddle{panel2.vps}{2cm}{0}{45}{35}{-0}{+75}
\plotfiddle{panel3.vps}{2cm}{0}{45}{35}{-250}{-100}
\plotfiddle{panel4.vps}{2cm}{0}{45}{35}{-0}{-25}
\caption{
\label{fig:1}
The excess-variance $\sigma^2_{rms}$ in the 0.5-10 keV band, 
from the combined SIS data, 
versus a) the rest-frame 2-10 keV luminosity in erg/s. 
b) $\Gamma_{2-10}$
c) FWHM H$\beta$ in km/s 
d) EW H$\beta$ in \AA\ . 
Sources with 
FWHM H$\beta < 2000$ km/s are shown as black squares, the rest 
of the sample are open crosses 
and error bars are 1 $\sigma$. }
\end{figure}

\end{document}